\documentclass[aps,pre,reprint,eqsecnum,showpacs]{revtex4-1}
\usepackage{amsmath}
\usepackage{graphicx}
\newcommand{\sech}{\hbox{sech}}
\newcommand{\csch}{\hbox{csch}}
\begin{document}

\title{Development of kinks in car-following models}
\date{\today}

\author{ Douglas A. Kurtze }
\affiliation{ Department of Physics, Saint Joseph's University, 5600 City Avenue, Philadelphia, PA 19131, USA }

\begin{abstract}

Many car-following models of traffic flow admit the possibility of absolute stability, a situation in which uniform traffic flow at any spacing is linearly stable.  Near the threshold of absolute stability, these models can often be reduced to a modified Korteweg-deVries (mKdV) equation plus small corrections.  The hyperbolic-tangent ``kink'' solutions of the mKdV equation are usually of particular interest, as they represent transition zones between regions of different traffic spacings.  Solvability analysis then shows that only a single member of the one-parameter family of kink solutions is preserved by the correction terms, and this is interpreted as a kind of selection.  We point out that one cannot extend the solvability analysis to a multiple-time-scales calculation, so that the solvability analysis does not point the way to any dynamical mechanism by which the ``selected'' kink might actually be selected.  On the other hand, we display a two-parameter family of traveling wave solutions of the mKdV equation which describe regions of one traffic spacing embedded in traffic of a different spacing; this family includes the kink solutions as a limiting case.  We carry out the multiple-time-scales calculation and find conditions under which the inclusions decay, conditions that lead to a selected inclusion, and conditions for which the inclusion evolves into a pair of kinks.  Finally, we show that the usual ``solvability'' calculation for kink solutions does not in fact identify a selected kink, but -- for all kink solutions -- merely gives a first-order correction to the relation between the traffic spacings far behind and far ahead of the kink.

\end{abstract}


\maketitle

\section{\label{sec:intro}Introduction}

Much progress in understanding the collective behavior of vehicular traffic has come from investigating car-following models, which describe the response of an individual vehicle to traffic conditions around it.  A typical car-following model specifies the acceleration of each car in terms of its current speed, the ``headway'' between it and the next car ahead of it, and the rate of change of the headway.  Many such models, such as that given in Eq. (\ref{startpt}) below, account for an explicit time delay between the traffic conditions and the response, so that the car's acceleration is determined by local conditions some finite time in the past.  Some incorporate further information, such as the positions and speeds of a finite number of cars ahead of and/or behind the car in question.  For reviews, see references \cite{Helbing-RMP2001}--\cite{LiSun-JCTA2012}.

Under quite general conditions \cite{Wilson-PTRSA2008}, these models have a continuous family of simple steady states that describe uniform traffic flow with an arbitrary constant spacing $\Delta$ between cars and all cars traveling at a constant speed that depends on $\Delta$.  Typically these steady states are linearly stable for some spacings and unstable for others, with instability occurring when the steady-state speed is too sensitive a function of $\Delta$.  Depending on the specifics of the acceleration function, it is also possible for {\it all\/} of these steady states to be linearly stable, a situation called ``absolute stability''.  (It is even possible to have ``absolute instability,'' with all steady states being linearly unstable, for example if the time delay is too long \cite{Kurtze-PRE2013}.)  Of course, it is of interest to go beyond linear stability analysis in order to understand what ultimately becomes of initially uniform traffic when its spacing is unstable, and also to uncover possible nonlinear instabilities \cite{OroszWilsonStepan-PTRSA2010}--\cite{OroszMoehlisBullo-PRE2010} of steady states that are {\it linearly\/} stable.

For steady states that are close to the onset of linear instability, one generally finds that, unless the time delay is long, the unstable perturbations to uniform flow are those in which the headway and speed vary slowly from car to car along the line of traffic, and they propagate upstream through the line of traffic at some finite phase velocity.  The stability analysis reveals how the relevant spatial and temporal scalings are related to the small deviation of $\Delta$ from the onset of instability, and one may then expand in powers of this deviation to reduce the car-following model to a nonlinear evolution equation for the headway, with the spatial independent variable being a scaled version of car number.  To leading order, this evolution equation turns out to be the Korteweg-deVries (KdV) equation.  In the case where the model is close to {\it absolute\/} stability, however, the relevant scalings are a bit different and one finds \cite{KomatsuSasa-PRE1995} a different leading-order equation, which in many cases (though not all \cite{Kurtze-PRE2013}, \cite{HayakawaNakanishi-PRE1998}) is the modified Korteweg-deVries (mKdV) equation.  This latter case is the focus of this paper.

Because the perturbations to uniform flow that grow near the onset of instability have long wavelengths, the derivations of the KdV and mKdV equations are quite robust to local changes in the model.  For example, if one allows the acceleration of a car to be affected by the second car ahead, or by the car behind, the derivations still go through with the new effects simply modifying the coefficients in the reduced equations.  Consequently, the mKdV equation has been derived near the threshold of absolute stability for a number of models \cite{GeChengDai-PhysA2005}--\cite{OuDaiDong-JPA2006}, including models with a second-neighbor effect \cite{MuramatsuNagatani-PRE1999} and with multiple-car look-ahead \cite{ShiChenXue-CTP2007}--\cite{TangHuangWongJiang-ActaMechSin2008}.  Similarly, near the onset of instability the linear growth rates of the growing perturbations are small, so the analysis is not sensitive to changes in the model that are local in time.  Thus the mKdV equation has also been derived for models with a finite time delay \cite{Nagatani-PRE1998}--\cite{NagataniNakanishi-PRE1998}, and even for models that include an anticipation effect, with a car's behavior depending on traffic conditions a short time in the future \cite{TianJiaLiGao-CPB2010}.  It bears repeating, however, that there is nonetheless a certain finite time delay which is large enough to make the initial instability occur at finite, rather than infinitely long, wavelengths \cite{Kurtze-PRE2013}, and that this would then lead to a different reduced equation.

Both the KdV and mKdV equations are well known to be exactly integrable, but usually one pays particular attention to specific families of traveling-wave solutions which propagate backward through the line of cars, namely the solitons for the KdV equation and the kinks for the mKdV equation.  A soliton would represent a localized concentration or rarefaction of traffic; a kink describes a localized transition between two different traffic spacings -- either the leading or trailing edge of a traffic jam, depending on whether the spacing behind the transition is smaller or larger than the spacing ahead of it.  In both cases these solutions form a one-parameter family, with the parameter determining the amplitude, width, and propagation rate of the solution.  Continuing the power-series expansion that led to the leading-order KdV or mKdV equation then yields first-order corrections to these evolution equations.  Typically, one then carries out a solvability analysis to determine the effect of these corrections, and finds a solvability condition which is satisfied by only a single parameter value.

For the mKdV kink solutions, this is a puzzling situation on the face of it.  The solvability result seems to indicate that only one of the kink solutions is the leading-order approximation to a solution of the full model.  However, the two traffic spacings that are connected by the kink are themselves given in terms of the kink parameter.  This means that there appears to be only {\it one\/} possible downstream spacing for which a kink solution is possible.  But the spacing far ahead of the transition zone is not an adjustable parameter -- it is set by the initial traffic spacing, before the occurrence of whatever perturbation led to the formation of the kink pattern.  Thus the solvability calculation says nothing about what happens when the initial traffic spacing is not equal to this special, ``selected'' value.  Moreover, it also says nothing about the dynamical process by which a kink pattern develops.

For a possible avenue toward answering these questions, we may look at the corresponding calculation for the one-soliton solutions of the KdV equation, which occurs when the traffic spacing is close to the onset of instability, but not near the {\it absolute\/} stability regime.  There also, a first-order perturbation calculation leads to a solvability condition which is satisfied by only one value of the soliton parameter.  One may extend this perturbation calculation, however, by allowing a slow time dependence of the kink parameter \cite{KurtzeHong-PRE1995}, \cite{ZhouLiuLuo-JPA2002}; the solvability condition is then replaced by an equation that describes how the soliton evolves along the family of solitons.  The soliton that satisfies the solvability condition is the fixed point of this evolution.  This multiple-time-scales approach also gives information about solitons that do not satisfy the solvability condition, and in particular it shows whether the ``selected'' soliton is a stable or unstable fixed point of the evolution.  In the latter case -- which turns out to be by far the more common -- the ``selected'' soliton will not be observed, as it merely marks the threshold of a nonlinear instability of uniform flow: solitons with smaller soliton parameter will decay, those with larger parameter value will grow.  For the mKdV kink solutions, however, we find that extending the perturbation calculation to allow a time evolution of the kink parameter does {\it not\/} help:  the coefficient of the time derivative of kink parameter turns out to be infinite \cite{Kurtze-PRE2013}.  In hindsight, this is not too much of a surprise, because adjusting the kink parameter entails changing the traffic spacing infinitely far behind and ahead of the transition zone, so it require moving individual cars by large distances.

The purpose of this paper is to investigate the role of the first-order corrections to the mKdV equation for traffic flow near absolute stability.  In particular, we wish to clarify the meaning of the ``selected'' kink -- the one that satisfies the solvability condition -- and to seek a dynamical mechanism by which a kink solution can develop from initially uniform, steady traffic.

In Section \ref{sec:derivation} we review how and when the mKdV equation arises when traffic conditions are near the threshold of absolute stability, establishing our notation as we do so.  In Section \ref{sec:inclusions} we display a {\it two}-parameter family of traveling-wave solutions of the leading-order mKdV equation which have equal upstream and downstream spacings, and so could arise in finite time from a localized perturbation to initially uniform traffic; we observe that the kink solutions are a nonuniform limit of this family.  We then investigate the effect of the correction terms in the equation on these solutions, finding that they drive a slow evolution of the second parameter.  In Section \ref{sec:diagram} we determine how this evolution plays out and what its final state will be, given the values of the relevant parameters and -- for some parameter ranges -- what the initial value of the parameter is.  In Section \ref{sec:kinks} we revisit the kink solutions, and explain the significance of the solvability calculation and of the ``selected'' kink parameter.  Finally, we summarize and discuss our results in Section \ref{sec:discussion}.

\section{\label{sec:derivation}Derivation of the mKdV equation}

We begin with a car-following model, which describes a single line of cars traveling along an infinitely long, uniform road, with all drivers behaving identically.  Such a model, of course, is most appropriate for describing cars controlled by on-board adaptive cruise control, since different human drivers will respond differently to identical road conditions, and indeed a given human driver will respond differently to identical conditions at different times.  We take the $x$ direction to be the direction of traffic flow and number the cars consecutively, with car $n+1$ ahead of car $n$.  The position of car $n$ at time $t$ is denoted by $x_n(t)$, and its velocity is $v_n = dx_n/dt$. The model is embodied in the equations of motion \cite{Wilson-PTRSA2008}
\begin{eqnarray}
  {dx_n(t) \over dt} &=& v_n(t), \nonumber \\
  {dv_n(t+t_d) \over dt} &=& A(x_{n+1}(t) - x_n(t), v_{n+1}(t) - v_n(t), v_n(t)).
    \label{startpt}
\end{eqnarray}
Here $t_d$ is a fixed delay time, so that car $n$ responds to conditions ahead of it as they were at a time $t_d$ before the present.  The acceleration function $A(h,\dot h,v)$ is a general function of the velocity $v = v_n$ of the car under consideration, the headway $h = x_{n+1}-x_n$ between it and the next car ahead, and the rate $\dot h = v_{n+1}-v_n$ at which the headway is changing.  For the model to be realistic, $A$ must be an {\it increasing\/} function of the headway and its rate of change and a {\it decreasing\/} function of the velocity of car $n$ itself \cite{Wilson-PTRSA2008}.

This model has a continuous family of steady states representing traffic flow with an arbitrary uniform spacing $\Delta$ between cars, given explicitly by
\begin{equation}
  x_{n+1} - x_n = \Delta \quad \hbox{and} \quad v_n = V_s(\Delta) \quad \hbox{for all $n$ and $t$},
    \label{uniformflow}
\end{equation}
where the steady-state traffic velocity $V_s(\Delta)$ is defined implicitly by
\begin{equation}
  A(\Delta,0,V_s(\Delta)) = 0.
    \label{def:Vs}
\end{equation}
If $A(h,0,v)$ is a decreasing function of $v$ (and does not remain constant for any finite range of $v$) then there is a unique $V_s$ for each possible spacing $\Delta$.

We will not present the linear stability analysis of these steady states here, but only quote the results relevant to our subsequent calculations.  For the model with {\it no\/} delay ($t_d=0$) the analysis is given by Wilson \cite{Wilson-PTRSA2008} and by Orosz et al. \cite{OroszWilsonStepan-PTRSA2010}; one finds that steady, uniform flow with spacing $\Delta$ is linearly unstable if the steady-state traffic speed $V_s(\Delta)$ is too sensitive to $\Delta$, specifically if
\begin{equation}
  V_s'(\Delta) > \Omega_c(\Delta) \equiv {1+2\lambda(\Delta)\over2\tau(\Delta)},
    \label{def:Omegac}
\end{equation}
where the prime denotes a derivative with respect to $\Delta$, and the parameters $\tau(\Delta)$ and $\lambda(\Delta)$ are given by
\begin{equation}
  {1 \over \tau} = -{\partial A \over \partial v}, \qquad
  {\lambda \over \tau} = {\partial A \over \partial \dot h},
    \label{def:lambda,tau}
\end{equation}
with the partial derivatives evaluated in the steady state $h = \Delta$, $\dot h = 0$, $v = V_s(\Delta)$.  Because $A(h,\dot h,v)$ must be an increasing function of $\dot h$ and a decreasing function of $v$, both $\tau$ and $\lambda$ are positive.  Note that $\lambda$ is dimensionless and $\tau$ has dimensions of time, so $\Omega_c$ has dimensions of inverse time (as does $V_s'$).  In addition to this stability criterion, one finds that just beyond the onset of instability, i.e., when the ``stability parameter'' $V_s'(\Delta) - \Omega_c(\Delta)$ is small and positive, it is the long-wavelength perturbations to uniform steady flow that grow.

For nonzero delay, the linear stability analysis is given by Orosz et al. \cite{OroszMoehlisBullo-PRE2010} and by Kurtze \cite{Kurtze-PRE2013}.  One finds that the above results for zero delay continue to hold \cite{Kurtze-PRE2013} if the time delay is not too large, specifically if $t_d$ is less than the smaller zero of the quadratic
\begin{equation}
  P = 1 - 2(1+\lambda)(t_d/\tau) + {1+2\lambda\over2}(t_d/\tau)^2.
    \label{def:P}
\end{equation}
For larger delay times the stability criterion is different, and uniform flow first becomes unstable to perturbations with finite wave number.  We will not consider that situation here.

It is entirely possible that for some realistic acceleration function, uniform steady flow turns out to be linearly stable for every spacing $\Delta$; this situation is called ``absolute stability''.  If the stability parameter $V_s'(\Delta) - \Omega_c(\Delta)$ has a maximum value of zero, then we say that the system is at the ``threshold'' of absolute stability; {\it near\/} threshold, then, the maximum value of the stability parameter is small.  If, in addition to the system being near the threshold of absolute stability, the traffic spacing $\Delta$ is close to {\it both\/} the maximum of the stability parameter {\it and\/} the inflection point $\Delta_i$ of $V_s(\Delta)$, then it is possible to reduce the model (\ref{startpt}) to a modified Korteweg-deVries (mKdV) equation, as we will show presently.  Of course, such a scenario can only occur if the maximum of the stability parameter lies close to $\Delta_i$.  However, many specific car-following models are defined in such a way that the stability threshold $\Omega_c$ is a constant independent of $\Delta$.  In such a model the maximum of the stability parameter automatically occurs {\it at\/} $\Delta_i$, and the reduction then goes through for any spacing near $\Delta_i$.  In general, however, $\Omega_c$ can be a nontrivial function of $\Delta$, and we do {\it not\/} obtain the mKdV equation unless the maxima of $V_s'$ (i.e., the inflection point) and of $V_s'-\Omega_c$ are at least close to one another \cite{Kurtze-PRE2013}.

Suppose now that the conditions above are satisfied:  the system is near absolute stability, so that the maximum value of $V_s' - \Omega_c$ is small, and this maximum occurs at a spacing $\Delta$ that is close to the inflection point $\Delta_i$ of $V_s$.  If the maximum value of $V_s' - \Omega_c$ is positive, then there is a range of linearly unstable spacings which is also small.  Thus we define an arbitrary, dimensionless small parameter $\epsilon$ so that the width of the unstable range is of order $\epsilon$.  Specifically, we write
\begin{equation}
  V_s'(\Delta_i) = \Omega_c(\Delta_i) + \epsilon^2 C \delta_i,
   \label{def:delta}
\end{equation}
where $C$ is a coefficient with dimensions of inverse time, to be chosen later.  The dimensionless parameter $\delta_i$, of order unity, characterizes how far from the threshold of absolute stability the system is.  In order for the maxima of $V_s'$ and $V_s' - \Omega_c$ to be within order $\epsilon$ of each other, we must also have $\Omega_c'$ small at $\Delta_i$, so we write
\begin{equation}
  \Omega_c'(\Delta_i) = \epsilon {C \over L} \omega_i,
   \label{def:omega}
\end{equation}
where $L$ is a coefficient with dimensions of length, also to be chosen later, and $\omega_i$ is dimensionless and of order unity.  For spacings within order $\epsilon$ of the inflection point, the stability parameter is then given by
\begin{widetext}
\begin{equation}
  V_s'(\Delta_i+\epsilon L \beta) - \Omega_c(\Delta_i+\epsilon L \beta) = \epsilon^2 C \left[ \delta_i - \omega_i\beta - {1\over2}{(\Omega_c''-V_s''')L^2 \over C}\beta^2 \right] + O(\epsilon^3),
\end{equation}
\end{widetext}
with the derivatives on the right side evaluated at $\Delta = \Delta_i$.  Note that $\Omega_c''-V_s'''$ must be positive in order for the extremum of $V_s'-\Omega_c$ to be a maximum; similarly, $V_s'''(\Delta_i)$ must be negative since $\Delta_i$ is a maximum of $V_s'$.  The stability parameter then has its maximum and its zeros, if any, for $\beta$ values of order unity, so they do in fact occur at spacings within order $\epsilon$ of $\Delta_i$.

Since the unstable modes near the onset of instability have small wave numbers and small linear growth rates, we expect that the {\it nonlinear\/} development of the flow near the onset of {\it linear\/} instability of the uniform steady state will take place on long spatial scales and slow time scales.  We then write the positions of the cars in the form
\begin{equation}
  x_n = n\Delta_i + V_s(\Delta_i)t + L f(z,T),
   \label{ansatz}
\end{equation}
where, motivated by the results \cite{Kurtze-PRE2013} of linear stability analysis with the above assumptions about $V_s'$ and $\Omega_c$, we define dimensionless scaled car-number and time variables
\begin{equation}
  z \equiv \epsilon [n + V_s'(\Delta_i) t], \qquad T \equiv \epsilon^3 C t.
   \label{def:z,T}
\end{equation}
Note that a given car is represented by a $z$ value that increases linearly with time, while a fixed function $f(z)$ represents a pattern of traffic that propagates through the line of cars against the direction of traffic at a rate (in cars per unit time) of $V_s'(\Delta_i)$.  Note also that the function
\begin{equation}
  g(z,T) \equiv {\partial f \over \partial z}
   \label{def:g}
\end{equation}
is then proportional, to leading order in $\epsilon$, to both the deviation in headway $x_{n+1}-x_n$ and the deviation in velocity $dx_n/dt$ of car $n$ from the steady state.

To derive the evolution equation for $f$, we substitute the expressions (\ref{def:delta}) and (\ref{def:omega}) and the ansatz (\ref{ansatz}) into the basic model equations (\ref{startpt}) and expand in powers of $\epsilon$.  The general expansion is given in Ref. \cite{Kurtze-PRE2013}; here it reduces to
\begin{widetext}
\begin{eqnarray}
 C f_T = C_{11} f_{zzz} + {1\over6} V_s''' L^2 f_z^3 + &\epsilon& \bigg[{1+2\lambda\over2}C(-\delta_i + \omega_i f_z) f_{zz}
  - {(1+2\lambda)^2\over8} P V_s' f_{zzzz} \nonumber \\*
  + {1+2\lambda\over4}(\Omega_c''-V_s''') L^2 f_z^2 f_{zz} &+& C_{22a} L f_z f_{zzz} + C_{22b} L f_{zz}^2 + {1\over24} V_s'''' L^3 f_z^4 \bigg] + \cdots,
 \label{expansion}
\end{eqnarray}
\end{widetext}
where subscripts on $f$ represent partial derivatives, and
\begin{subequations}
\begin{eqnarray}
 C_{11} &=& \left[{1+3\lambda \over 6} - {(1+2\lambda)^2\over4} {t_d\over\tau} \right] V_s', \label{C11} \\
 C_{22a} &=& \left({\lambda'\over2} - \tau' t_d V_s'^2 \right) V_s', \\
 C_{22b} &=& {1\over8} V_s' \left({\partial A\over\partial v^2} - 4 {\partial A\over\partial\dot h\partial v} + 4 {\partial A\over\partial\dot h^2}\right),
\end{eqnarray}
\end{subequations}
with the partial derivatives of $A$ again evaluated in the steady state and all $\Delta$-dependent parameters and their derivatives evaluated at $\Delta=\Delta_i$.  We note in passing that if we regard the right side of (\ref{C11}) as a function of $\Delta$, take its derivative, and evaluate at $\Delta_i$, the result is $C_{22a}$.

While the function $f$ gives the deviation between the actual position of car $n$ and the position it would have in the exact uniform-flow steady state, the deviations of speed and headway are usually of more direct interest, so we differentiate (\ref{expansion}) with respect to $z$ and write the result in terms of $g=f_z$.  We also choose $C = C_{11}$ and $L^2 = 12C/|V_s'''|$ for convenience.  This gives us
\begin{widetext}
\begin{eqnarray}
 g_T = g_{zzz} - 2(g^3)_z + \epsilon {\partial\over\partial z} \bigg[{1+2\lambda\over2}(&-&\delta_i + \omega_i g) g_z  - {(1+2\lambda)^2 P V_s' \over 8C_{11}} g_{zzz} \nonumber \\*
  + {1+2\lambda\over4}{(\Omega_c''-V_s''') L^2 \over C_{11}} g^2 g_z &+& {C_{22a}L \over C_{11}} g g_{zz} + {C_{22b}L \over C_{11}} g_z^2 + {V_s'''' L^3 \over 24C_{11}} g^4 \bigg] + \cdots,
 \label{mKdVplus}
\end{eqnarray}
\end{widetext}
which is the mKdV equation plus correction terms.  Note that, as a result of the arbitrariness of the expansion parameter $\epsilon$, the entire equation is invariant under the rescaling $\epsilon \to k\epsilon$, $g \to g/k$, $z \to kz$, $T \to k^3T$, $\delta \to \delta/k^2$, $\omega \to \omega/k$ for any $k$.

Although we chose the underlying spacing in the expansion to be $\Delta_i$, this entails no loss of generality.  One may verify, by a straightforward but lengthy calculation, that the final equation for $g$ is the same for any other basic spacing that is within order $\epsilon$ of $\Delta_i$.

\section{\label{sec:inclusions}Inclusions and their development}

The leading order of the reduced evolution equation (\ref{mKdVplus}) is the defocusing modified Korteweg-deVries (mKdV) equation, which is exactly integrable via an inverse scattering procedure.  Here we will only concern ourselves with traveling wave solutions of the form $g(z-uT)$, where -- unlike the familiar hyperbolic-tangent ``kink'' solutions -- $g$ approaches the {\it same\/} constant value $g_\infty$ for both $z \to -\infty$ and $z \to \infty$.  Such a solution could develop in finite time from a localized perturbation to initially uniform, steady traffic; it represents traffic of a fixed background spacing $\Delta_i + \epsilon L g_\infty$ with an ``inclusion'' propagating upstream at a rate (again, in {\it cars\/} per unit time) of $V_s' - \epsilon^2 C_{11} u$, where both $u$ and the shape $g$ of the inclusion are to be determined by the calculation.  For these solutions, the leading order of (\ref{mKdVplus}) becomes
\begin{equation}
  0 = g_{zzz} - 2(g^3)_z + u g_z.
  \label{mKdV}
\end{equation}
Integrating once introduces an arbitrary constant of integration and gives an equation for $g$ which is analogous to the Newtonian equation of motion of a particle in a quartic potential, with $z$ and $g$ playing the roles of, respectively, time and the particle's position.  The quartic term in the potential is negative, so in order to have solutions in which $g$ remains bounded, the two arbitrary parameters -- $u$ and the constant of integration -- must be chosen so that the potential has two local maxima.  If the energy of the analogue particle is lower than the height of both maxima, we obtain solutions for $g$ that are periodic in $z$, which we will not consider here.  If the two maxima have equal height and the particle energy is equal to that height, then we obtain the one-parameter family of kink solutions.  We will focus here on the solutions that occur when the maxima have unequal heights and the particle energy is equal to the height of the lower maximum.  There is a two-parameter family of these solutions, because for any $u$ in some range, there is a range of values of the constant of integration for which the potential has unequal maxima.  A straightforward calculation yields the explicit form of these solutions:
\begin{subequations}
\begin{eqnarray}
  g^{(0)}(z) &=& g_\infty \left( 1 - {2 \sinh\alpha \tanh\alpha \over \cosh kz + \cosh\alpha} \right) \label{inclusion} \\
             &=& g_\infty \left[ 1 - \tanh\alpha \left(\tanh {kz+\alpha\over2} - \tanh {kz-\alpha\over2} \right) \right],
   \label{inclusion2}
\end{eqnarray}
\end{subequations}
with
\begin{equation}
  k = 2 |g_\infty| \tanh\alpha, \qquad u = 2 g_\infty^2 (1 + 2\,\sech^2\alpha).
\end{equation}
The parameter $g_\infty$, as in the kink solutions, can be positive or negative; its arbitrariness reflects the scaling invariance of the mKdV equation, which in turn arises from the arbitrariness of the expansion parameter $\epsilon$.  The second parameter, $\alpha$, which can be taken to be nonnegative, controls the shape of the inclusion and its propagation rate.  Inclusions with larger $\alpha$ have a larger disparity between the spacing in the inclusion and the background spacing, and propagate slightly more rapidly through the line of traffic.  From (\ref{inclusion}) we see that an inclusion with small $\alpha$ is a small, broad deviation from uniform flow, with the minimum value of $g/g_\infty$ being approximately $1-\alpha^2$ and the width of order $k^{-1}\sim\alpha^{-1}$.  From (\ref{inclusion2}) we see that an inclusion with large $\alpha$ is a plateau (at a value close to $-g_\infty$) between two hyperbolic-tangent kinks separated in $z$ by $2\alpha/k \approx \alpha/|g_\infty|$.  In particular, if we redefine $z = \alpha/k$ to be the origin, then the kink solution is the (nonuniform) limit of inclusion solutions for $\alpha\to\infty$.  Thus the inclusion solutions interpolate smoothly between uniform traffic flow and a kink/anti-kink pair.

To see the effect of the order-$\epsilon$ correction terms in (\ref{mKdVplus}), we carry out a standard multiple-time-scales calculation.  We write $g$ in the form
\begin{equation}
  g(z,T) = g^{(0)}(z-uT;\alpha(\epsilon T)) + \epsilon g^{(1)}(z-uT,T) + \cdots,
  \label{gexpansion}
\end{equation}
thus allowing the inclusion parameter $\alpha$ to vary slowly with time.  Note that this form splits the first-order correction into two parts, one involving a change in $\alpha$ and one not.  At an appropriate point in the calculation, we will need to impose some subsidiary condition to define precisely how this split is made.  Leaving that condition unspecified for the time being, we substitute (\ref{gexpansion}) into (\ref{mKdVplus}) and expand to first order in $\epsilon$ to obtain
\begin{equation}
 {\partial g^{(0)} \over \partial \alpha} \dot\alpha + {\partial g^{(1)} \over \partial T} = {\partial^3 g^{(1)} \over \partial z^3} - 6 {\partial \over \partial z} g^{(0)2} g^{(1)} + u {\partial g^{(1)} \over \partial z} + {\partial\over\partial z} [\cdots],
 \label{g1equation}
\end{equation}
where the overdot denotes differentiation with respect to the new slow time $\epsilon T$, and the square brackets contain all the terms in the square brackets in (\ref{mKdVplus}) with $g$ replaced by $g^{(0)}$.

To isolate an equation for $\dot\alpha$, we multiply both sides of this equation by $g_\infty - g^{(0)}$, which vanishes for $z\to\pm\infty$, and integrate over all $z$.  Since we are taking the initial condition to be uniform steady traffic plus a {\it localized\/} perturbation, the full solution $g$ must continue to approach $g_\infty$ at $z\to\pm\infty$, and so the correction $g^{(1)}$ must vanish at $\pm\infty$.  Integration by parts then eliminates the terms on the right side involving $g^{(1)}$.  We now choose the subsidiary condition mentioned above to be
\begin{equation}
 \int_{-\infty}^{\infty} (g_\infty - g^{(0)}) {\partial g^{(1)} \over \partial T} \, dz = 0,
\end{equation}
so that {\it all\/} terms involving $g^{(1)}$ drop out of the equation for $\dot\alpha$.  The remaining integrals can be evaluated analytically; since $g^{(0)}$ is an even function, the last three terms in the square brackets do not contribute.  Thus we obtain our evolution equation for $\alpha$:
\begin{widetext}
\begin{equation}
 \dot\alpha = {1+2\lambda\over3} g_\infty^2 \left\{ f_0(\alpha) \tilde\delta + f_1(\alpha) \left[ \tilde\omega g_\infty - 5 {(1+2\lambda) P V_s' \over 4C_{11}} g_\infty^2 \right] - f_2(\alpha) \left[ \tilde\kappa g_\infty^2 - 7 {(1+2\lambda) P V_s' \over 4C_{11}} g_\infty^2 \right] \right\}.
  \label{alphaeqn}
\end{equation}
The new parameters
\begin{eqnarray}
 \tilde\delta &=& \delta_i - \omega_i L g_\infty - {1\over2} {(\Omega_c'' - V_s''') L^2 \over C_{11}} g_\infty^2, \nonumber\\*
 \tilde\omega &=& \omega_i L + {(\Omega_c'' - V_s''') L^2 \over C_{11}} g_\infty, \nonumber\\*
 \tilde\kappa &=& {(\Omega_c'' - V_s''') L^2 \over C_{11}}
\end{eqnarray}
are scaled values of the stability parameter $V_s'-\Omega_c$ and its negative $\Delta$-derivatives evaluated at the traffic spacing $\Delta_i + \epsilon Lg_\infty$ far from the inclusion rather than at the inflection point $\Delta_i$ of the steady traffic speed $V_s$.  The $\alpha$-dependent coefficients are given by
\begin{eqnarray}
 f_0(\alpha) &=& 3\coth\alpha - 2\tanh\alpha - 3\alpha\csch^2\alpha, \nonumber\\*
 f_1(\alpha) &=& {1\over2} [15\coth\alpha - 13\tanh\alpha - 3\alpha\csch^2\alpha(5 - \tanh^2\alpha)], \nonumber\\*
 f_2(\alpha) &=& {1\over10} [105\coth\alpha - 115\tanh\alpha + 16\tanh^3\alpha - 15\alpha\csch^2\alpha(7 - 3\tanh^2\alpha)].
\end{eqnarray}
These are all positive functions, which approach constant values of 1, 1, and 3/5 respectively for large $\alpha$ and vanish as $2\alpha^3/5$, $8\alpha^5/35$, and  $8\alpha^7/105$ for $\alpha\to0$.

For small $\alpha$, $f_1$ and $f_2$ are much smaller than $f_0$, so (\ref{alphaeqn}) reduces to
\begin{equation}
 \dot\alpha = \left[ {2(1+2\lambda)\over15} g_\infty^2 \tilde\delta \right] \alpha^3 + O(\alpha^5).
  \label{smallalphaeqn}
\end{equation}
An inclusion with small $\alpha$ is a small deviation from steady, uniform traffic flow with a spacing of $\Delta_i + \epsilon Lg_\infty$.  For negative $\tilde\delta$ this steady state is linearly stable, and we see that $\alpha$ does indeed decay (as $T^{-1/2}$), so the inclusion reverts to uniform, steady flow.  On the other hand, for positive $\tilde\delta$ the inclusion grows.

For large $\alpha$, (\ref{alphaeqn}) reduces to
\begin{equation}
 \dot\alpha = {1+2\lambda\over3} g_\infty^2 \left[ \tilde\delta + \tilde\omega g_\infty - {3\over5} \tilde\kappa g_\infty^2  - {(1+2\lambda) P V_s' \over 5C_{11}} g_\infty^2 \right] + O(\alpha e^{-2\alpha}).
 \label{largealphaeqn}
\end{equation}
\end{widetext}
Thus when $\alpha$ is large it increases or decreases {\it linearly\/} with $T$, according to whether the quantity in brackets is positive or negative.  For large $\alpha$ the inclusion is a pair of kinks separated by a $z$ range of $\alpha/|g_\infty|$, so in this regime the number of cars in the inclusion grows or shrinks linearly with time.  Since $z$ is defined relative to a reference frame which moves backward through the line of traffic, this amounts to having the leading and trailing edges of the inclusion moving backward relative to traffic at slightly different rates.

It is also possible, of course, for (\ref{alphaeqn}) to admit stable or unstable fixed points.  We now examine the parameter ranges in which the various possible long-time behaviors occur.

\section{\label{sec:diagram}Long-time behavior}

According to (\ref{alphaeqn}), the fixed-point structure of the slow evolution of inclusions is governed by three parameters, as we may see by factoring out the (positive) quantity $(1+2\lambda)PV_s'g_\infty^2/4C_{11}$ to obtain
\begin{equation}
 \dot\alpha = {(1+2\lambda)^2\over12C_{11}} P V_s' g_\infty^4 [f_0(\alpha) \delta + f_1(\alpha) (\omega - 5) - f_2(\alpha) (\kappa - 7)].
 \label{scaledeqn}
\end{equation}
The new parameters here are scaled versions of the stability parameter and its negative derivatives at the far-downstream traffic spacing,
\begin{eqnarray}
 \delta &=& {4\tilde\delta \over (1+2\lambda) P V_s' g_\infty^2}, \nonumber\\*
 \omega &=& {4\tilde\omega \over (1+2\lambda) P V_s' g_\infty}, \nonumber\\*
 \kappa &=& {4\tilde\kappa \over (1+2\lambda) P V_s'}.
\end{eqnarray}
Note again that because the system is near a {\it maximum\/} of the stability parameter, $\kappa$ must be positive.

A straightforward approach, calculating the value(s) of $\alpha$ for which $\dot\alpha$ vanishes, is analytically impractical because of the complicated form of the functions $f_i$.  Instead, we will regard $\alpha$ as an independent variable and look for the loci in parameter space where the evolution has a fixed point -- whether stable or unstable -- for a given value of $\alpha$.  From (\ref{scaledeqn}) we see that this occurs for
\begin{equation}
 \omega = 5 - {f_0(\alpha) \over f_1(\alpha)} \delta + {f_2(\alpha) \over f_1(\alpha)} (\kappa - 7).
\end{equation}
For a given $\delta$, this is a line in the $\kappa$-$\omega$ plane with slope $f_2(\alpha)/f_1(\alpha)$ and $\omega$-intercept $[5f_1(\alpha) - 7f_2(\alpha) - \delta f_0(\alpha)]/f_1(\alpha)$.  As we increase $\alpha$ from 0 to $\infty$, the slope increases monotonically from zero to 3/5, the $\delta$-independent term in the intercept decreases monotonically from 5 to 4/5, and the coefficient of $\delta$ in the intercept increases monotonically from $-\infty$ to $-1$.  The $\alpha\to\infty$ line,
\begin{equation}
 \omega = (4 - 5\delta + 3\kappa)/5,
\end{equation}
is of particular importance:  from (\ref{scaledeqn}) we see that for $\omega$ above this line, $\dot\alpha$ approaches a positive constant for large $\alpha$, so that a sufficiently large inclusion will grow linearly with time, eventually replacing the initial traffic spacing, $\Delta_i + \epsilon Lg_\infty$, with a new spacing $\Delta_i - \epsilon Lg_\infty$.  It is perhaps interesting to note that this new spacing would have a $\delta$ value which is given in terms of the parameters for the original spacing by $\delta + 2\omega - 2\kappa$, and an $\omega$ value of $-\omega + 2\kappa$.  Parameters for the {\it new\/} spacing would then also be above the $\alpha\to\infty$ line, since this transformation leaves $5\delta + 5\omega - 3\kappa$ unchanged, so that the new spacing would also be susceptible to the formation of inclusions that would then reestablish the old spacing.

\begin{figure}[b]
\includegraphics [width=8.6 cm, keepaspectratio=true]{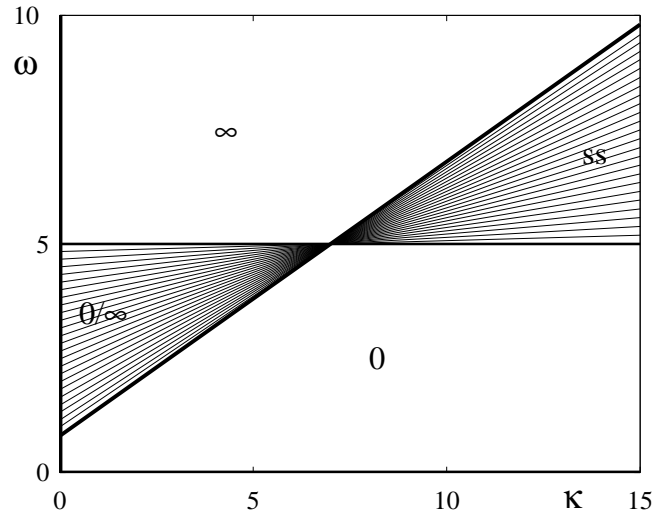}
\caption{\label{figure1} Fixed-point lines in the $(\kappa,\omega,\delta)$ parameter space with $\delta=0$.  Along each line the inclusion evolution equation (\ref{scaledeqn}) has a fixed point at a given $\alpha$, ranging from $\alpha=0$ on the heavy horizontal line $\omega=5$ to the $\alpha\to\infty$ limit on the heavy diagonal line $\omega=(4+3\kappa)/5$.  For parameters in the region marked ``0'' any initial inclusion decays to $\alpha=0$; in the ``$\infty$'' region any initial inclusion grows toward $\alpha=\infty$; in the ``ss'' region any initial inclusion evolves toward a steady state with the fixed-point $\alpha$.  In the ``0/$\infty$'' region the fixed point is unstable, so an initial inclusion with $\alpha$ below the fixed-point value decays to $\alpha=0$, while one with $\alpha$ above it grows toward $\alpha=\infty$.}
\end{figure}

The analysis is simplest for the case when the initial traffic spacing is marginally stable, so that $\delta=0$.  It is then trivial to see that all the fixed-point lines pass through the point $\kappa=7$, $\omega=5$.  These lines are plotted in Figure 1; note that they divide the $\kappa$-$\omega$ plane into four wedges.  For parameters in the upper left wedge, where $\omega$ is greater than 5 and also above the $\alpha\to\infty$ line, $\dot\alpha$ is positive for all $\alpha$ -- it is positive for large $\alpha$ because parameters are above the $\alpha\to\infty$ line, and it has no zeros -- so any initial inclusion will grow to infinite $\alpha$.  Similarly, in the lower right wedge, where $\omega$ is less than 5 and below the $\alpha\to\infty$ line, $\dot\alpha$ is negative for all $\alpha$, so that any initial inclusion will decay back to the uniform state.  In the upper right wedge, where $\omega$ is above 5 but below the $\alpha\to\infty$ line, $\dot\alpha$ is positive for small $\alpha$, changes sign at the fixed-point $\alpha$ value, and remains negative for large $\alpha$.  The fixed point is then stable, since $\alpha$ increases if below it and decreases if above it.  In this wedge any initial inclusion approaches a fixed point, with the fixed-point $\alpha$ getting higher as we move higher in the wedge.  Finally, in the lower left wedge, with $\omega$ below 5 but above the $\alpha\to\infty$ line, $\dot\alpha$ is negative for small $\alpha$ and positive for large $\alpha$, changing sign at the fixed-point value.  Thus the fixed point is unstable, and marks the threshold of a finite-amplitude instability: any initial inclusion with $\alpha$ below the fixed-point value will decay back to $\alpha=0$, while one with $\alpha$ above the fixed-point value with grow to $\alpha\to\infty$.  The threshold gets smaller as we move upward through the wedge, from the region where $\alpha=0$ is the final state for all initial inclusions toward the region where all initial inclusions go to $\alpha\to\infty$.

\begin{figure}[b]
\includegraphics [width=8.6 cm, keepaspectratio=true]{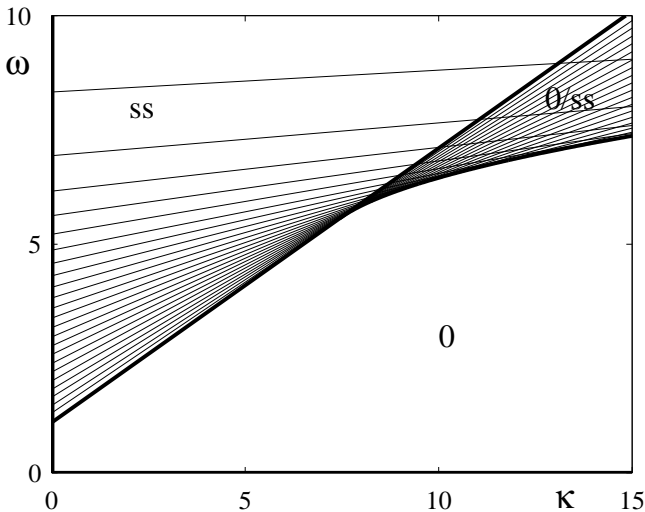}
\caption{\label{figure2} As in Fig. 1, but with $\delta = -0.3$, typical for negative $\delta$.  The $\omega$-intercept of the fixed-point lines decreases and the slope increases with increasing $\alpha$.  For parameters in the overlap region marked ``0/ss'' there are two fixed-point values of $\alpha$.  An initial inclusion with $\alpha$ below the smaller of the two decays to $\alpha=0$, while one with $\alpha$ larger than the smaller fixed-point value evolves toward the larger fixed-point $\alpha$.}
\end{figure}

For $\delta<0$, so that the background spacing $\Delta_i + \epsilon Lg_\infty$ is linearly stable, a new possibility arises.  The fixed-point lines for a typical case are plotted in Figure 2.  As for $\delta=0$, the slope increases and the intercept decreases monotonically with increasing $\alpha$, but the fixed-point lines no longer all pass through a single point. Instead there is now an overlap region in which pairs of fixed-point lines cross.  In addition, for small $\alpha$ the intercept goes as $\alpha^{-2}$, so fixed-point lines cover the plane up to arbitrarily large $\omega$.  There are now three distinct regions of the parameter plane.  Everywhere above the $\alpha\to\infty$ line, $\dot\alpha$ is negative for small $\alpha$ and positive for large $\alpha$, crossing zero at the single fixed-point $\alpha$ value.  As in the lower left wedge for $\delta=0$, this is a region of bistability, in which an initial inclusion decays to zero if its $\alpha$ is below the (unstable) fixed-point value and grows to $\alpha\to\infty$ if above it.  Below the $\alpha\to\infty$ line is a region with no fixed point, in which any initial inclusion decays to zero.  In the overlap region, which is also below the $\alpha\to\infty$ line, $\dot\alpha$ is negative for small $\alpha$, turns positive at the smaller fixed-point $\alpha$, then turns negative again at the larger one and remains negative for large $\alpha$.  The larger-$\alpha$ fixed point is then stable; any initial inclusion with $\alpha$ above the {\it smaller\/} fixed-point value approaches it at long time, while any with $\alpha$ smaller than the smaller fixed-point value decays to zero.  At the boundary between the overlap region and the region without fixed points, then, the two fixed-point values coincide and $\dot\alpha$ has a double zero.  At this double zero, both $\dot\alpha$ as given in (\ref{scaledeqn}) and its $\alpha$-derivative are equal to zero.  Solving these two equations for $\kappa$ and $\omega$ yields parametric equations for the boundary of the overlap region, i.e. the envelope of the fixed-point lines:
\begin{eqnarray}
 \omega - 5 &=& -\left[{d\over d\alpha}{f_0(\alpha)\over f_2(\alpha)}/{d\over d\alpha}{f_1(\alpha)\over f_2(\alpha)}\right] \delta, \nonumber \\
 \kappa - 7 &=& \left[{d\over d\alpha}{f_0(\alpha)\over f_1(\alpha)}/{d\over d\alpha}{f_2(\alpha)\over f_1(\alpha)}\right] \delta.
  \label{junction}
\end{eqnarray}
We find that for small $\alpha$ this curve approaches the inverse parabola $\omega \approx (-7\kappa\delta/3)^{1/2}$ with large $\kappa$, while for large $\alpha$ it meets the $\alpha\to\infty$ line tangentially at $\kappa = 7 - (5/4)\delta$, $\omega = 5 - (7/4)\delta$.

\begin{figure}[b]
\includegraphics [width=8.6 cm, keepaspectratio=true]{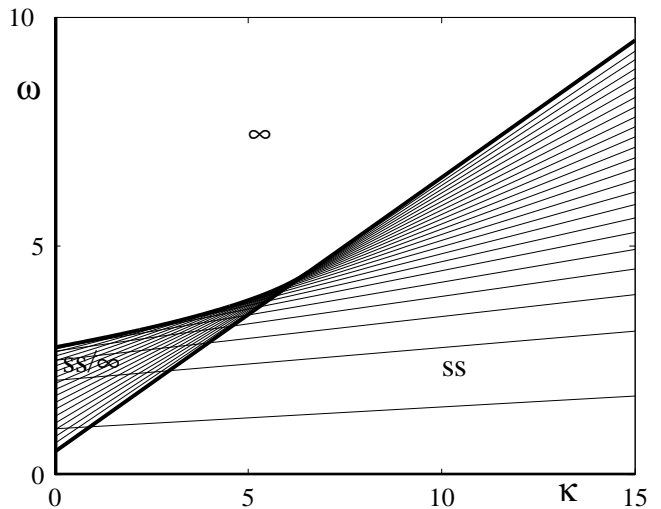}
\caption{\label{figure3} As in Figs. 1 and 2 but for $\delta = 0.3$, typical for positive $\delta$ less than 28/5.  For parameters in the ``ss'' region any initial inclusion evolves toward the fixed-point $\alpha$.  In the ``ss/$\infty$'' region there are two fixed-point $\alpha$ values; an initial inclusion with $\alpha$ below the larger of the two evolves toward the smaller, while one with $\alpha$ above the larger fixed-point value grows toward $\alpha=\infty$.  This region is absent for $\delta>28/5$.}
\end{figure}

For positive $\delta$ the background spacing is linearly unstable; the fixed-point lines for a typical case, with $\delta$ not too large, are plotted in Figure 3.  The $\omega$-intercept of the fixed-point lines now approaches $\omega \to -\infty$ for $\alpha \to 0$, and it increases with increasing $\alpha$ until -- for $\delta$ not too large -- it reaches a maximum.  Thus the fixed-point lines initially fan upward.  After the intercept reaches its maximum, further fixed-point lines form an overlap region, as in the $\delta<0$ case except now lying above the $\alpha\to\infty$ line.  Below the $\alpha\to\infty$ line, $\dot\alpha$ is positive for small $\alpha$ and turns negative at the fixed-point value of $\alpha$.  In this region, then, the fixed point is stable and any initial inclusion will tend toward it.  Above the overlap region and the $\alpha\to\infty$ line, $\dot\alpha$ is positive for all $\alpha$, so any initial inclusion grows toward $\alpha\to\infty$.  In the overlap region, $\dot\alpha$ is positive for both small and large $\alpha$ and negative between the two fixed-point values, so this is again a bistable region: an initial inclusion with $\alpha$ below the larger fixed-point value tends toward the smaller fixed-point $\alpha$, and one with $\alpha$ above the larger fixed point tends to infinity.  As in the negative-$\delta$ case, the upper boundary of the overlap region is where $\dot\alpha$ has a local minimum at a value of zero, and so the parametric equations for that boundary are the same as (\ref{junction}).  As above, the boundary merges with the $\alpha\to\infty$ line at $\kappa = 7 - (5/4)\delta$, $\omega = 5 - (7/4)\delta$.  For $\delta>28/5$ this junction would occur at negative $\kappa$, so the intercept never reaches a maximum as $\alpha$ increases, and there is no overlap region.

\begin{figure}[b]
\includegraphics [width=8.6 cm, keepaspectratio=true]{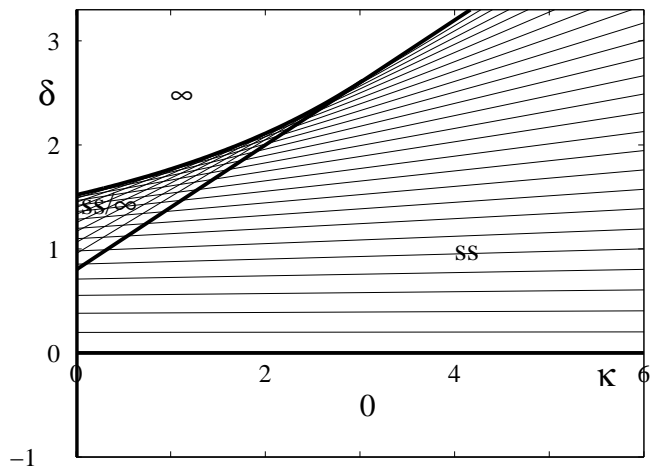}
\caption{\label{figure4} Fixed-point lines for $\omega = 0$.  The slope of the fixed-point lines increases with $\alpha$, from $\alpha=0$ on the bold horizontal line $\delta=0$ to $\alpha\to\infty$ on the heavy diagonal line $\delta=(4+3\kappa)/5$.  Labeling of the regions is the same as for Figs. 1-3.  The upper boundary of the ``ss/$\infty$'' region meets the vertical axis at $\delta=1.51881$ and merges with the diagonal line at $\delta=20/7$, $\kappa=24/7$.}
\end{figure}

The most unstable spacing has $\omega=0$; the fixed-point lines for this case are plotted on the $\kappa$-$\delta$ plane in Figure 4.  Note that if the ``most unstable'' spacing is in fact linearly stable, so that $\delta<0$, then all initial inclusions decay to $\alpha=0$, so that there is no finite-amplitude instability in this case.  If it is actually linearly {\it un}stable, then an initial inclusion approaches a steady-state inclusion if $\delta$ is less than $(4+3\kappa)/5$, while for larger $\delta$ it is possible for an initial inclusion to grow to infinity, again replacing the initial traffic spacing $\Delta_i + \epsilon L g_\infty$ with a new spacing $\Delta_i - \epsilon L g_\infty$.  For intermediate $\delta$ with $\kappa$ not too large, there is a region of bistability, where the $\alpha$ value of the initial inclusion determines whether the final state will be a stable inclusion or growth to $\alpha\to\infty$.

\section{\label{sec:kinks}Kink solutions and solvability}

To find the familiar kink solutions of the mKdV equation, we return to the mechanics analogue in Section \ref{sec:inclusions}, choose parameters so that the two maxima of the quartic ``potential'' are at equal heights, and choose the ``energy'' to be that common maximum value.  In this way we obtain
\begin{equation}
  g^{(0)}(z-uT) = g_\infty \tanh |g_\infty|(z-uT)
   \label{kink}
\end{equation}
with
\begin{equation}
  \qquad u = 2 g_\infty^2,
\end{equation}
where $g_\infty$ is a free parameter.  We can now try to repeat the multiple-time-scales calculation to find an equation for the slow evolution of the kink parameter $g_\infty$ that is driven by the corrections to the mKdV equation.  This fails, however, because when we multiply the analogue of (\ref{g1equation}) by $g^{(0)}$ and integrate over all $z$, the coefficient of $\dot g_\infty$ diverges.  This procedure, then, does {\it not\/} uncover any dynamical mechanism that could select one of the infinite family of kink solutions.  What, then, is the significance of the solvability condition that must be satisfied in order to find a first-order correction $g^{(1)}$?

This quandary becomes yet more puzzling if we think of the pattern of traffic as having arisen from a localized perturbation to initially uniform, steady flow.  The model equations (\ref{startpt}) contain no mechanism for downstream traffic to respond to conditions upstream, i.e. for larger $n$ (or $z$) to be affected by smaller $n$ (or $z$).  Thus the traffic spacing far downstream, at $z\to+\infty$, must always remain equal to the initial spacing; this becomes a boundary condition on the reduced equation (\ref{mKdVplus}).  This boundary condition, in turn, fixes the value of $g_\infty$ immediately.  But the spacing far downstream is $\Delta_i + \epsilon L g_\infty$, and since there is no reason why traffic must have started at the particular spacing for which $g_\infty$ satisfies the solvability condition, we almost always have a paradoxical situation where the boundary condition forbids satisfying the solvability condition.  What then becomes of initially steady, uniform traffic at a spacing that does not satisfy solvability?

To see how to resolve this situation, we retrace the steps that lead to the solvability condition.  First we write the perturbed spacing as
\begin{equation}
  g(z,T) = g^{(0)}(z-uT) + \epsilon g^{(1)}(z-uT) + \cdots
\end{equation}
with
\begin{equation}
  \qquad u = 2 g_\infty^2 + \epsilon u^{(1)},
\end{equation}
thus allowing an order-$\epsilon$ correction to the propagation rate $u$.  We substitute this expansion into (\ref{mKdVplus}) and linearize in $\epsilon$ to obtain the analogue of (\ref{g1equation}),
\begin{equation}
 0 = {\partial \over \partial z} \left\{ {\partial^2 g^{(1)} \over \partial z^2} - 6 g^{(0)2} g^{(1)} + 2g_\infty^2 g^{(1)} + u^{(1)} g^{(0)} + [\cdots] \right\}.
 \label{kinkcorrection}
\end{equation}
Next we multiply this equation by $g^{(0)}$ and integrate by parts.  Crucially, in order to eliminate the terms involving $g^{(1)}$, we must impose the boundary conditions $g^{(1)} \to 0$ at $z \to \pm\infty$.  However, while the boundary condition at large {\it positive\/} $z$ is clearly appropriate, because the {\it downstream\/} spacing is fixed at $\Delta_i + \epsilon L g_\infty$, there is no compelling reason to demand that the far-{\it upstream\/} density be precisely $\Delta_i - \epsilon L g_\infty$.  Consequently the appropriate boundary condition for $z \to -\infty$ is just that $g^{(1)}$ approach some undetermined constant.  With this less restrictive boundary condition, the integrations by parts now yield boundary terms, and instead of a solvability condition we obtain a formula for the upstream spacing,
\begin{widetext}
\begin{eqnarray}
  g^{(1)}(-\infty) = {1 \over C_{11}} \bigg\{ {1+2\lambda\over 6} \bigg[ &\delta&_i - {1+2\lambda\over 5}PV_s'g_\infty^2 - {1\over 10}(\Omega_c''-V_s''') L^2 g_\infty^2 \bigg] \hbox{ sgn } g_\infty \nonumber \\*
  + {2\over15}(&C&_{22a}-2C_{22b})Lg_\infty^2 + {1\over60}V_s''''L^3g_\infty^2 \bigg\}.
\end{eqnarray}
\end{widetext}
The right side of this equation is the quantity which, according to the solvability condition, is supposed to vanish.  Thus the traffic spacing that satisfies the conventional solvability condition is not one that is somehow ``selected,'' but rather it is merely the initial spacing for which the average of the upstream and downstream spacings remains $\Delta_i$ through first order in $\epsilon$.

We may also determine the correction to the propagation rate of the kink by simply integrating (\ref{kinkcorrection}) over all $z$; the result is
\begin{equation}
  u^{(1)} = -2 g_\infty g^{(1)}(-\infty).
\end{equation}

We can also obtain these results from the inclusion calculation, by multiplying (\ref{g1equation}) by $g^{(0)}$ and integrating only from $0$ to $\infty$ instead of from $-\infty$ to $\infty$.  When $\alpha$ is large, according to (\ref{inclusion2}), for positive $z$ the inclusion solution differs from the kink solution only by corrections that are exponentially small in $\alpha$.  Moreover, $\dot\alpha/2$ is the first-order shift in velocity of the kink.

\section{\label{sec:discussion}Discussion}

A common calculation for car-following models, when parameter values are near the threshold of absolute stability, is to reduce the model to a modified Korteweg-deVries (mKdV) equation with small corrections, write down the one-parameter family of hyperbolic-tangent ``kink'' solutions of the mKdV equation, and then carry out a solvability analysis to find which one of the kink solutions -- the ``selected'' kink -- persists when the correction terms are included.  The significance of this selected kink, however, has always been rather problematic.  The calculation gives no hint as to how the dynamics of the model might actually select it, nor does it give any hint of what might happen if the traffic spacing far ahead of or far behind the kink does not match the value in the selected kink solution.  For parameter values near the onset of instability, rather than near {\it absolute\/} stability, a similar sequence of calculations is common:  one reduces the model to a Korteweg-deVries (KdV) equation with small corrections, writes down the one-parameter family of one-soliton solutions, and carries out a solvability analysis to find the selected soliton.  In this situation it is possible to extend the solvability calculation to a multiple-time-scales analysis, which shows that the correction terms drive a slow evolution along the family of soliton solutions, with the ``selected'' soliton being the fixed point of this evolution.  This analysis further shows whether the soliton parameter evolves toward or away from its ``selected'' value; in the former case, that soliton is in fact dynamically selected, while in the latter it marks a finite-amplitude instability threshold, with smaller initial soliton parameters decaying to zero and larger ones growing to infinity (and into a regime in which the assumptions underlying the derivation of the KdV equation are no longer valid).  It is crucial, however, that the soliton solutions are localized while the kink solutions are not.  We have found that the corresponding multiple-time-scales analysis for kink solutions fails: there is no slow evolution along the family of kink solutions.  This is perhaps not particularly surprising, since different kink solutions have different traffic spacings far ahead of and far behind the kink, so that any small change to the kink parameter requires changing those spacings, which then entails large changes to the {\it positions\/} of cars infinitely far ahead of and far behind the kink.  Moreover, most car-following models include no mechanism for cars to adjust their behavior to traffic conditions {\it behind\/} them, so the car spacing infinitely far ahead of the kink must be fixed.  This means, however, that the spacing far ahead of the kink is ultimately set by initial conditions, not by any solvability consideration -- a fact which makes the difficulties noted above even more consequential.

This last observation points the way toward understanding the actual significance of the ``selected'' kink.  The conventional solvability calculation implicitly assumes that the first-order correction to the kink solution vanishes far ahead of the kink and also far behind it.  The underlying model, however, does not treat those two regions symmetrically.  The development of the traffic pattern proceeds backwards from some localized perturbation to initially uniform traffic, so that the traffic spacing far ahead of the developing kink is fixed, but the spacing behind it is not.  It is reasonable, then, to choose the zeroth-order kink solution to be the one which has the correct {\it downstream\/} spacing, but there is no reason why this zeroth-order solution should also get the {\it upstream\/} spacing right to anything beyond leading order.  This implies that one must allow the first-order correction to approach a nonzero constant, not necessarily zero, far behind the kink.  If we modify the solvability calculation to allow this, we find that it gives only an equation for the correction to the spacing far behind the kink.  The ``selected'' kink is merely the one for which this first-order correction vanishes.

To investigate how a kink solution could arise from initially uniform, steady traffic flow, we have examined the two-kink, or ``inclusion,'' solutions of the mKdV equation.  For a given initial spacing, there is a one-parameter family of these solutions, interpolating between a small, broad perturbation to uniform flow and a widely separated kink-antikink pair.  Corrections to the mKdV equation then cause a slow time evolution of the parameter, and there are three possibilities for the long-time result of this evolution:  an initial inclusion might decay to zero, or it may evolve to a stable inclusion at some selected parameter value, or it may develop into a kink-antikink pair, with the kink and antikink propagating through the line of traffic at slightly different rates, so that the region between them grows linearly with time.  Which of these scenarios ultimately plays out depends on three parameters, namely the parameter governing linear stability of uniform traffic at the initial spacing and its first and second derivatives with respect to spacing.  We have delineated the regions of parameter space in which each possible long-time behavior occurs; there are some regions in which more than one long-time behavior is possible, and initial conditions determine which actually occurs.  In those situations in which the inclusion develops into a widely separated kink-antikink pair, we find that the kink and the antikink are exponentially close to the familiar kink solutions.

It is worth noting that a large inclusion essentially replaces the initial spacing by one that is as far below the inflection point as the initial one was above it (plus corrections).  This is a somewhat unsatisfactory aspect of our results, as we would like to see solutions in which jams form which replace the initial spacing with other spacings as well.  This would allow the possibility of jam formation leading to some traffic spacing which actually {\it is\/} selected.  In connection with this idea, however, we note that when parameters are in a range in which the initial spacing can be replaced by a growing inclusion with the new spacing, then the new spacing could also be replaced by a growing inclusion of the old spacing.  Thus it may be that the tendency is simply to produce traffic patterns in which regions of traffic at different spacings alternate irregularly, rather than to converge to some selected spacing everywhere.

The fact that our inclusion solutions allow only a single possibility for the spacing within the inclusion seems to arise from our having looked for leading-order solutions which are stationary in some reference frame, i.e. for which the leading and trailing edges of the inclusion move through the line of traffic at the same rate.  In future we will seek and examine more general inclusion solutions which are not stationary (analogous to those found for the Korteweg-deVries equation by the Hirota method \cite{Dodd-book1982}) to see whether a more comprehensive picture emerges of the effect of the correction terms on these solutions.  Taking this idea further, it is known that the mKdV can be solved for arbitrary initial conditions, at least in principle, by an inverse scattering transform; one could try to use this as a basis for a perturbation theory by writing the corrections in terms of the scattering variables \cite{KarpmanMaslov-PL1977}.  While complicated, this approach would offer the possibility of identifying globally attracting traffic patterns.

Finally, it would be of interest to extend these results, and other analytic results for car-following models, to models with longer time delay.  As we have noted, our results are valid provided the explicit time delay in the model is small enough that the parameter $P$ defined in (\ref{def:P}) is positive, and $P$ itself then appears in one of the correction terms to the mKdV equation.  Dealing with a larger time delay, however, is not simply a matter of allowing $P$ to be negative.  For larger delays the initial instability does not occur at infinitely long wavelengths, so that the basis of the reduction of the model to an mKdV equation is not valid, and it is necessary to restart the analysis from this much earlier point.

\end{document}